\documentclass[epjST]{svjour}
\usepackage{graphicx}
\usepackage{amsfonts}
\usepackage{amsmath}
\begin{document}
\title{Advantages of $q$-logarithm representation over 
$q$-exponential representation from
the sense of scale and shift on nonlinear systems}
\author{Hiroki Suyari\inst{1}\fnmsep\thanks{\email{suyari@faculty.chiba-u.jp}} \and Hiroshi Matsuzoe\inst{2} \and Antonio M. Scarfone\inst{3}}
\institute{Graduate School of Engineering, Chiba University, 1-33, Yayoi-cho, Inage-ku, Chiba 263-8522, Japan \and Graduate School of Engineering, 
Nagoya Institute of Technology, Gokiso-cho, Showa-ku, Nagoya 466-8555, 
Japan \and Istituto dei
Sistemi Complessi (ISC-CNR) c/o, Politecnico di Torino, Corso Duca degli
Abruzzi 24, Torino I-10129, Italy}
\abstract{
Addition and subtraction of observed values can be computed under the obvious and implicit assumption that the scale unit of measurement should be the same for all arguments, which is valid even for any nonlinear systems.
This paper starts with the distinction between exponential and 
non-exponential family in the sense of the scale unit of measurement.
In the simplest nonlinear model ${dy}/{dx}=y^{q}$, it is shown how typical effects such as rescaling and shift emerge in the nonlinear systems and affect observed data. Based on the present results, the two representations, namely the $q$-exponential and the $q$-logarithm ones, are proposed.
The former is for rescaling, the latter for unified understanding with a fixed scale unit.
As applications of these representations, the corresponding entropy and the general probability expression
for unified understanding with a fixed scale unit are presented.
For the theoretical study of nonlinear systems, $q$-logarithm representation is shown to have significant advantages over $q$-exponential representation.
} 
\maketitle
%
\section{Introduction}
\label{intro}
In Boltzmann-Gibbs-Shannon theory, i.e., the standard statistical mechanics 
\cite{To10}\cite{Gr10} and information theory \cite{CT91}, most of the
important probability distributions such as canonical distribution, Gaussian
distribution, and probability for optimal code length belong to the
so-called exponential family \cite{NG09}. The distributions in the
exponential family follow the exponential law: 
\begin{equation}
\exp\left( x\right) \exp\left( a\right) =\exp\left( {x+a}\right),\quad
\exp\left( x\right) /\exp\left( a\right) =\exp\left( {x-a}\right) 
\label{explaw_plus}
\end{equation}
which play significant roles in every computation within this family.
This law represents the
operation by the \textit{shift} in each argument, which means that
multiplication and division in the exponential family is just given by plus
and minus \textit{shift} in arguments, respectively:
\begin{equation}
x\mapsto x+a,\quad x\mapsto x-a.   \label{shift}
\end{equation}
On the other hand, if we consider a power-law distribution out of the
exponential family, such \textit{shift} operations in multiplication and
division disappear: 
\begin{equation}
x^{-\gamma}a^{-\gamma} =\left( {xa}\right) ^{-\gamma}, \quad
x^{-\gamma}/a^{-\gamma} =\left( {x/a}\right) ^{-\gamma}.
\end{equation}
Instead, \textit{rescaling} is emerging:
\begin{equation}
x\mapsto xa,\quad x\mapsto x/a.   \label{rescaling}
\end{equation}

Let us compare shift (\ref{shift}) and rescaling (\ref{rescaling}) from the
sense of the scale unit of measurement in the following example. 
Consider a situation in which there are two rulers with two different
scale units to measure a length on $\mathbb{R}$ (see Fig.1). 
\begin{figure}[htbp]
\vspace{-0.7cm}
\hspace{1.5cm}A given length:
\setlength\unitlength{1truecm}
\begin{picture}(3,1)
\put(0,0){\line(1,0){3}}
\multiput(0,0)(3,0){2}{\line(0,1){0.3}}
\end{picture}
\par\vspace{-1mm}\hspace{2.46cm} Ruler 1:
\setlength\unitlength{1truecm}
\begin{picture}(3,1)
\put(0,0){\line(1,0){6}}
\multiput(0,0)(1,0){6}{\line(0,1){0.7}}
\multiput(0,0)(0.5,0){12}{\line(0,1){0.5}}
\multiput(0,0)(0.1,0){58}{\line(0,1){0.3}}
\end{picture}
\par\vspace{-4mm}\hspace{2.46cm} Ruler 2:
\setlength\unitlength{1.5truecm}
\begin{picture}(2,1)
\put(0,0){\line(1,0){4}}
\multiput(0,0)(1,0){4}{\line(0,1){0.5}}
\multiput(0,0)(0.5,0){8}{\line(0,1){0.3}}
\multiput(0,0)(0.1,0){38}{\line(0,1){0.2}}
\end{picture}
\caption{Ruler1 and ruler 2 with different scale unit length}\label{fig:01}
\end{figure}
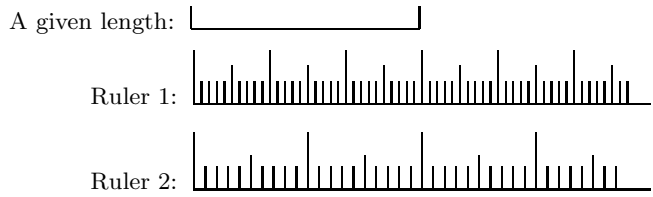

For a given length, one ruler (ruler 1) indicates 3 meters and 
the other (ruler 2) 2 meters when we measure it 
with these two different rulers.
In this example, the units of measurement (e.g., meter) are the same, 
but the scale units of measurement of these two rulers are different 
from each other.
Of course, if we use the correct ruler, we obtain the correct length. 
However, the correct scale unit of measurement is determined by humans, 
and nature does \textit{not} depend on kinds of rulers.
Then, in the shift (\ref{shift}) $x$ and $a$ \textit{must} have 
the same scale units of measurement, so that the computations 
$x+a$ and $x-a$ can be done. 
Thus the scale unit of measurement must be invariant
over addition and subtraction.
In the exponential family, any multiplication and division can be done under the obvious invariance of the scale unit of measurement in any argument. But in the rescaling (\ref{rescaling}) the scale unit of measurement is \textit{not} invariant if $a\neq1$.

In general, scale units variant observation can be found in nonlinear dynamics
with rescaling.
Especially, in sequential observations, each observation ideally should have the same scale unit of measurement to deal with data in science or engineering. 
Thus, the assumption of independence among observations is the
most ideal, which does not yield scale change in each observation. The
invariance of scale unit of measurement is captured 
by the functions such as the probability distributions
in the exponential family.
However, some correlations due to rescaling
can often be observed in nonlinear systems, 
which leads to one of the reasons for the emergence of power-law distributions
far from exponential one.

In order to find a unified understanding of these two operations
 (shift (\ref{shift}) and rescaling (\ref{rescaling})) in the simplest way, 
we go back to the foundation: the simplest nonlinear generalization 
characterizing the exponential function
\begin{equation}
\frac{{dy}}{{dx}}=y^{q}.   \label{nlde}
\end{equation}
The choice of the starting point (\ref{nlde}) in the present work originates
from two aspects: statistical physics and mathematics.
In statistical physics, especially for generalization of Boltzmann-Gibbs statistics, (\ref{nlde}) is the basis for sensitivity to initial conditions, relaxation time, and stationary state (see \cite{Ts04} for details).
In the mathematical sense, (\ref{nlde}) recovers the famous characterization of
 $\exp\left( x\right)$ for the shift (\ref{shift}) when $q\rightarrow1$.
Moreover, (\ref{nlde}) is expected to have the rescaling (\ref{rescaling}) due to the nonlinearity when $q\neq1$.


\section{Scale unit of measurement in the nonlinear systems}
\subsection{Scale unit of measurement, inevitably determined by the initial
condition}
Obviously, (\ref{nlde}) is a nonlinear differential equation 
with respect to $y$.
But, if the following generalized logarithm, the so-called $q$%
-logarithm defined by 
\begin{equation}
\ln_{q}y:=\int_{1}^{y}{\frac{1}{{v^{q}}}}dv=\frac{{y^{1-q}-1}}{{1-q}} 
\label{lnq}
\end{equation}
is employed, (\ref{nlde}) is reformed to a \textit{linear} differential
equation with respect to $\ln_{q}y$. 
\begin{equation}
\frac{{d\ln_{q}y}}{{dx}}=1\quad\text{i.e.,}\quad \ln_{q}y=x+\ln_{q}C_{0}.
\label{dnlde}
\end{equation}
Here $C_{0}$ is a positive real number determined by 
\begin{equation}
\ln_{q}C_{0}=\ln_{q}y_{0}-x_{0}   \label{initial condition}
\end{equation}
for an initial condition $\left( {x_{0},y_{0}\left( >0\right) }\right) $ in
(\ref{nlde}).

Equation (\ref{dnlde}) is reformed to 
\begin{equation}
\frac{y}{C_{0}}=\exp_{q}\left( {\frac{x}{{C_{0}^{1-q}}}}\right),
\label{reform_sol}
\end{equation}
where 
\begin{equation}
\exp_{q}x:=\left[ {1+\left( {1-q}\right) x}\right] ^{\frac{1}{{1-q}}} 
\label{expq}
\end{equation}
for $1+\left( {1-q}\right) x>0$, which is the inverse function of $\ln_{q}x$ 
and is called $q$-exponential function.


Therefore, for the rescaling: 
\begin{equation}
\tilde{y}:=\frac{y}{C_{0}},\quad\tilde{x}:=\frac{x}{{C_{0}^{1-q}}}, 
\label{rescaling2}
\end{equation}
(\ref{reform_sol}) is rewritten as 
\begin{equation}
{\tilde{y}}=\exp_{q}\left( {\tilde{x}}\right) .
\end{equation}
This means that the nonlinear differential equation (\ref{nlde}) is \textit{%
invariant} under the rescaling (\ref{rescaling2}), i.e., 
\begin{equation}
\frac{{d\tilde{y}}}{{d\tilde{x}}}=\tilde{y}^{q}.   \label{rnlde}
\end{equation}

\begin{proposition}[rescaling]
The nonlinear differential equation (\ref{nlde}) is invariant under the rescaling (\ref%
{rescaling2}).
\end{proposition}

The rescaling factor $C_{0}$ is determined by an initial condition $\left( {%
x_{0},y_{0}\left( >0\right) }\right) $ in (\ref{nlde}) with $%
\ln_{q}C_{0}=\ln_{q}y_{0}-x_{0}$ (see (\ref{initial condition})), which
implies that $C_{0}$ can be taken as \textit{any} positive real number. In
other words, an initial condition 
$\left( {x_{0},y_{0}\left( >0\right) }\right) $ determines
the scale unit of measurement in (\ref{nlde}).

Then, in (\ref{reform_sol}) the elementary scale unit \textquotedblleft$1$%
\textquotedblright\ of observed value appears as a unit in the argument of
the $q$-exponential function $\exp_{q}$ such that ${x}/{{C_{0}^{1-q}}}=1$%
, i.e., $x={C_{0}^{1-q}}$. For a different initial condition $\left( {%
x_{1},y_{1}\left( >0\right) }\right) $ with ${x_{0}\neq x_{1}}$ and $%
\ln_{q}C_{1}=\ln_{q}y_{1}-x_{1}$, $x={C_{1}^{1-q}}$ is similarly obtained as
its elementary scale unit \textquotedblleft$1$\textquotedblright\ of
observed value. 
When $q=1$, the elementary scale unit \textquotedblleft$1$%
\textquotedblright\ of observed value always appears as $x=1$ (of
course!) which does \textit{not} depend on the initial condition of the
corresponding differential equation.
However, as shown above, in the
nonlinear dynamics governed by (\ref{nlde}), the scale unit of observed value
inevitably depends on the initial condition.
Therefore, when $q\neq1$, the usual normalization for probability 
depends on the scaling effect
on observed value ($x$-axis), so that the normalization in the case $q\neq1$ 
should be very careful, as discussed in detail 
in the last section.

\begin{figure}[h]
\begin{center}
\includegraphics[scale=0.7]{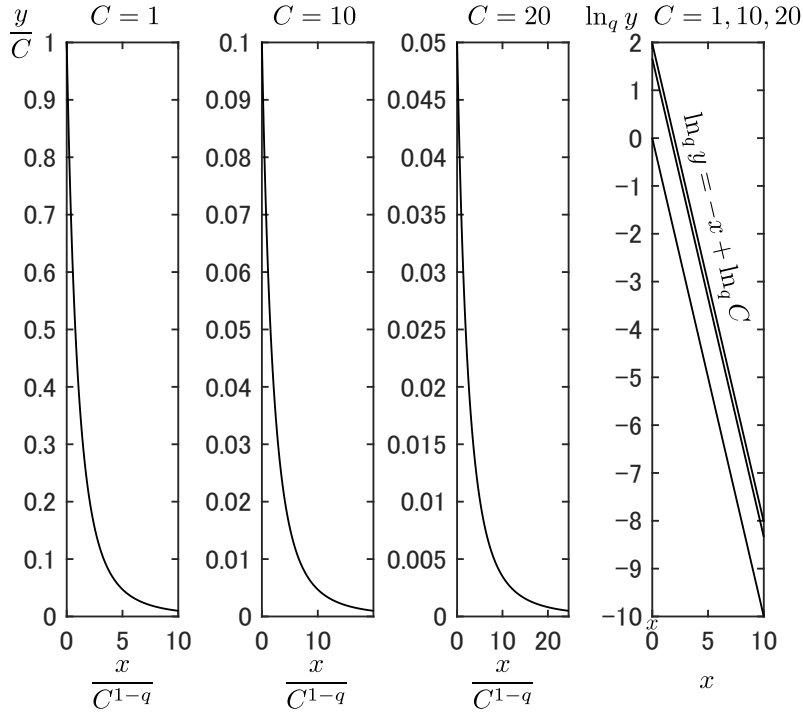}%
\caption{$\dfrac{y}{C}=\exp_{q}\left(-\dfrac{x}{C^{1-q}}\right)$
for $C=1,10,20$ (the left three graphs) and 
$\ln_{q}y=-x+\ln_{q}C$ (the rightmost graph)
where $q=1.3$.}
\end{center}
\end{figure}

For example, graphs of $\dfrac{y}{C}=\exp_{q}\left(  -\dfrac{x}{C^{1-q}%
}\right) $ for $C=1,10,20$ are described
in the left three figures of Fig.2.
The shapes of the left three graphs in Fig.2 are completely the same,
but each scale unit of measurement is different with each other
due to the rescaling $C$ (see both $x$-axis and $y$-axis
in the left three graphs of Fig.2).
This means under the rescaling (\ref{rescaling2}) the graph of
$y=\exp_{q}\left( -x\right)$ is invariant.
The invariance under the rescaling (\ref{rescaling2}) is confirmed
by the same slope of $\ln_{q}y=-x+\ln_{q}C$
(see the rightmost graph in Fig.2).

\subsection{\label{subsection for shift}Scale unit of measurement,
inevitably changed by shift}

In the previous subsection, for a given nonlinear differential equation (\ref%
{nlde}) a rescaling (\ref{rescaling2}) in both $x$ and $y$ arguments
inevitably appears. More precisely, the nonlinear differential
equation (\ref{nlde}) is \textit{invariant} under the rescaling (\ref%
{rescaling2}) (see (\ref{nlde}), (\ref{rescaling2}), and (\ref{rnlde})).
Such a rescaling can appear without using the nonlinear differential
equation (\ref{nlde}), that is, \textit{shift} in argument. For a given $%
y=\exp_{q}\left( x\right) $, if we apply a shift $x\mapsto x+c$ to this
equation, we obtain%
\begin{equation}
y=\exp_{q}\left( x+c\right) =\exp_{q}\left( c\right) \exp_{q}\left( \frac{x}{%
\left( \exp_{q}\left( c\right) \right) ^{1-q}}\right) , 
\label{expansion by shift}
\end{equation}
that is, 
\begin{equation}
\frac{y}{\exp_{q}\left( c\right) }=\exp_{q}\left( {\frac{x}{\left(
\exp_{q}\left( c\right) \right) ^{1-q}}}\right) .   \label{reform_sol2}
\end{equation}
Thus, by the rescaling: 
\begin{equation}
y^{\prime}:=\frac{y}{\exp_{q}\left( c\right) },\quad x^{\prime}
:=\frac{x}{\left(
\exp_{q}\left( c\right) \right) ^{1-q}},   \label{rescaling3}
\end{equation}
we obtain 
\begin{equation}
y^{\prime}=\exp_{q}\left( x^{\prime}\right) .   \label{result of shift}
\end{equation}
This means that $y=\exp_{q}\left( x\right) $ is \textit{invariant} under a
shift $x\mapsto x+c$ in argument $x$, which yields the same rescaling as (%
\ref{rescaling2}).

These two operations rescaling and shift in $y=\exp_{q}\left( x\right) $ are
equivalent to each other. In fact, for a given rescaling such as (\ref%
{rescaling3}) we obtain $y=\exp_{q}\left( x+c\right) $ which is a shift $%
x\mapsto x+c$ in $x$-argument of $y=\exp_{q}\left( x\right) $. On the other
hand, for a given shift such as (\ref{expansion by shift}) we can get a
rescaling (\ref{rescaling3}).

\begin{proposition}[shift and rescaling]
A shift $x\mapsto x+c$ to $y=\exp_{q}\left( x\right) $
for any $c\in \mathbb{R}$
satisfying $1+\left( 1-q\right) c>0$ is equivalent to a rescaling in both $x$%
-axis and $y$-axis.
\end{proposition}

Shift in the argument of the $q$-exponential function results in various scale
units of measurement in sequential observations. According to the property
of the $q$-exponential function: 
\begin{equation}
\exp_{q}\left( x_{1}+\cdots+x_{n}\right) =\exp_{q}\left( x_{1}\right)
\cdots\exp_{q}\left( \frac{x_{n}}{1+\left( 1-q\right) \sum _{i=1}^{n-1}x_{i}}%
\right) \text{,}   \label{property1}
\end{equation}
$x_{1},\cdots,x_{n}$ on the left side must have a same scale unit of
measurement, so that the sum $x_{1}+\cdots+x_{n}$ can be computed.
On the other hand, we get the observed values $x^{\prime}_{1},\cdots,x^{\prime}
_{n}$ on $\mathbb{R}$ (i.e., $\left( x^{\prime}_{1},\cdots
,x^{\prime}_{n}\right) \in\mathbb{R}^{n}$) with different scale units of
measurement such as%
\begin{equation}
x_{1}^{\prime}=x_{1},\quad x_{2}^{\prime}=\frac{x_{2}}{1+\left( 1-q\right)
x_{1}},\quad\cdots,\quad x_{n}^{\prime}=\frac{x_{n}}{1+\left( 1-q\right)
\sum_{i=1}^{n-1}x_{i}}.   \label{def of x_prime}
\end{equation}
Recall that $x_{1},\cdots,x_{n}$ have the same scale unit, so that
observed values $x^{\prime}_{1},\cdots,x^{\prime}_{n}$ have different
scale units if $q\neq1$. This representation is due to the property of the $q
$-exponential (\ref{expq}).

As shown in the study of the dynamics determined by (\ref{nlde}), 
there exist two representations, namely the $q$-exponential representation and 
the $q$-logarithm representation. The choice of these
two representations depends on what we want to express. $q$-Exponential
representation is useful for rescaling, while $q$-logarithm representation for
unified studies with a fixed scale unit of measurement.

Note that \lq\lq {\it unified studies}" in the $q$-logarithm representations
mean that it is possible to study the dynamics
with a {\it fixed} scale unit of measurement.
On the other hand, in the $q$-exponential representations such as (\ref{property1}), observed values $x^{\prime}_{1},\cdots,x^{\prime}_{n}$
in (\ref{def of x_prime}) have {\it different} scale unit of measurement, which makes the unified studies difficult in general.

\section{Two representations in the systems determined by the fundamental
nonlinear differential equation}

\subsection{$q$-Exponential representation for rescaling}

If we want to represent a rescaling effect in our formulations, $q$%
-exponential representation such as (\ref{reform_sol}) 
and (\ref{property1}) is more useful than
the corresponding $q$-logarithm representation given in (\ref{dnlde}).
In fact, $q$-exponential representation reveals how each variable in the
formulation is rescaled by other variables or constants 
(e.g., (\ref{property1})).

But there are some disadvantages to using $q$-exponential representation.
One of these is the appearance of complicated rescaling in sequential
observations. For a given $q$-exponential representation $y=\exp_{q}\left(
x\right) $, a shift in $x$ such that $x\mapsto x+c_{1}$ is applied to this $q
$-exponential representation. Then, in the same way as (\ref{expansion by
shift}) we obtain%
\begin{equation}
\frac{y}{\exp_{q}\left( c_{1}\right) }=\exp_{q}\left( \frac{x}{\left(
\exp_{q}\left( c_{1}\right) \right) ^{1-q}}\right) \text{.} 
\label{expansion by shift2}
\end{equation}
Again, one more shift in the argument of $q$-exponential function is applied
to this expression (\ref{expansion by shift2}), then we can get%
\begin{equation}
\frac{y}{\exp_{q}\left( c_{1}\right) \cdot\exp_{q}\left( c_{2}\right) }%
=\exp_{q}\left( \frac{x}{\left( \exp_{q}\left( c_{1}\right) \right)
^{1-q}\left( \exp_{q}\left( c_{2}\right) \right) ^{1-q}}\right) . 
\label{2steps rescaling}
\end{equation}
Note that a shift by $c_{1}$ is different from that by $c_{2}$ in the sense
of scale unit. More concretely, a shift by $c_{1}$ is given by $x\mapsto
x+c_{1}$, but a shift by $c_{2}$ is given by ${x}/{\left( \exp_{q}\left(
c_{1}\right) \right) ^{1-q}}\mapsto{x}/{\left( \exp_{q}\left(
c_{1}\right) \right) ^{1-q}}+c_{2}$. Then, scale unit of shift $c_{1}$ is
different from that of $c_{2}$.

Here we need to make some comments on the $q$-product \cite{NMW03}\cite{Bo03}. 
As discussed in the
previous section, $x_{1},\cdots,x_{n}$ on the left side of (\ref%
{property1}) must have the same scale unit of the measurement, 
but for the observed values $x^{\prime}_{1},\cdots,x^{\prime}_{n}$ 
appeared on the right side does
not so. In particular, each $x^{\prime}_{t}$ has different scale unit 
of the measurement by rescaling with past internal values (often called \lq\lq state variables" in control theory) $x_{t-1},x_{t-2},\cdots $, 
which makes theoretical analysis difficult. In order to avoid these
difficulties, the $q$-product is useful in many applications \cite{Su04b}%
\cite{Ts09}\cite{Sc13}. The $q$-product $\otimes_{q}$ is introduced to
satisfy 
\begin{equation}
\exp_{q}\left( x_{1}+x_{2}\right) =\exp_{q}\left( x_{1}\right) \otimes
_{q}\exp_{q}\left( x_{2}\right)   \label{q-exponential law}
\end{equation}
as a generalization of the exponential law \cite{NMW03}\cite{Bo03}. Then,
the property (\ref{property1}) can be rewritten by means of the $q$-product. 
\begin{equation}
\exp_{q}\left( x_{1}+x_{2}+\cdots+x_{n}\right) =\exp_{q}\left( x_{1}\right)
\otimes_{q}\exp_{q}\left( x_{2}\right)
\otimes_{q}\cdots\otimes_{q}\exp_{q}\left( x_{n}\right).
\label{representation by q-product}
\end{equation}
Therefore, the $q$-product \textit{preserves} scale unit of measurement 
among $x_{1},\cdots,x_{n}$, 
so that there are a lot of successful applications in
this field \cite{Su04b}. But at the same time, there are some disadvantages
to use the $q$-product as shown below.

One of some disadvantages using the $q$-product is as follows: From the
requirement (\ref{q-exponential law}), the definition of the $q$-product $%
\otimes_{q}$ is given by%
\begin{equation}
x\otimes_{q}y:=\left[ x^{1-q}+y^{1-q}-1\right] ^{\frac{1}{1-q}},
\label{q-product}
\end{equation}
which is valid only under the constraints $x,y>0$ and $x^{1-q}+y^{1-q}-1>0$.
In each computation by means of $q$-product or $q$-ratio (inverse operation
of the $q$-product), it should be confirmed if these constraints are 
satisfied or not. 
Another disadvantage is that there is no room to employ a scaling effect $%
C$ in the formulations using the $q$-product. Of course, a scaling effect $C$
can be added in ad hoc such that\ $y=\exp_{q}\left( x\right)
\otimes_{q}\exp_{q}\left( C\right) $, but this expression does not show a
rescaling effect in arguments.

\subsection{$q$-Logarithm representation for unified studies with a fixed scale
unit of measurement}

As shown in (\ref{reform_sol}) and (\ref{rescaling2}), a scaling factor $%
C_{0}$ (i.e., initial condition) affects significantly on observed data 
in the nonlinear dynamics.
In the dynamics governed by the fundamental nonlinear differential equation (%
\ref{nlde}) the scaling factor $C_{0}$ is determined by the initial
condition (\ref{initial condition}) and inevitably appears in (\ref%
{dnlde}) or (\ref{reform_sol}). If $q$-exponential representation is
used in formulations such as (\ref{reform_sol}) and (\ref{property1}), a
scaling factor $C_{0}$ appears in every argument (e.g., both sides in (\ref%
{reform_sol}) and $x_{1}\left( =\ln_{q}C_{0}\right) $ on the right side of (%
\ref{property1})).
This strong dependence of $C_{0}$ on each argument yields
serious difficulties in analysis and understanding.
However, in $q$-logarithm representation such as (\ref{dnlde}) 
(the origin of (\ref{reform_sol})),
a scaling factor $C_{0}$ appears only one time in one formula
which has a lot of advantages over $q$-exponential representation. 
For example, in (\ref{dnlde}), a shift in $x$ like $x\mapsto x+c$ is 
described by just a shift of a graph on a $x$-$q$-log plot.

Moreover, in $q$-logarithm representation such as (\ref{dnlde}), all arguments
have the same scale unit of measurement. 
On the other hand, in the $q$-exponential representation (\ref{reform_sol}), 
{scale units of }$x$ and ${x}/{{C_{0}^{1-q}}}$ are obviously 
different with each other. Thus, $q$-logarithm
representation has an important advantage over $q$-exponential
representation in the sense of scale unit.


\section{Application of $q$-logarithm representation}
\subsection{Rederivation of Tsallis entropy via $q$-logarithm representation}

In \cite{Su04b}, $q$-product (\ref{q-product}) is applied to the
derivation of Tsallis entropy as the unique entropy corresponding to the
fundamental nonlinear differential equation (\ref{nlde}). For the following
discussion, let us briefly review how several formulations such as $q$%
-Stirling's formula and Tsallis entropy can be uniquely obtained from the
fundamental nonlinear differential equation (\ref{nlde}) with some
modifications of the original version \cite{Su04b}. The distinction from the
original derivation is that the $q$-product is \textit{not} explicitly used
to avoid some difficulties stated in the previous section.

For any natural number $n\in\mathbb{N}$, the $q$-logarithm of the $q$%
-factorial is introduced:%
\begin{equation}
\ln_{q}n!_{q}:=\sum_{k=1}^{n}\ln_{q}k.
\label{q-logarithm of the q-factorial}
\end{equation}
Then, for large $n\in\mathbb{N}$ we can get the $q$-Stirling's formula: 
\begin{equation}
\ln_{q}n!_{q}\simeq\left\{ 
\begin{array}{l}
\dfrac{n}{2-q}\ln_{q}n-\dfrac{n}{2-q}+\dfrac{1}{2}\ln_{q}n+\dfrac{1}{2-q}%
\quad(q\neq2) \\ 
n-\ln n-\dfrac{1}{2n}-\dfrac{1}{2}\text{\quad}\quad\qquad\qquad\qquad
\quad\left( q=2\right)%
\end{array}
\right. .   \label{q-Stirling's formula}
\end{equation}
By means of (\ref{q-logarithm of the q-factorial}), the $q$-logarithm of the 
$q$-multinomial coefficient is defined by 
\begin{equation}
\ln_{q}\left[ 
\begin{array}{ccc}
& n &  \\ 
n_{1} & \cdots & n_{k}%
\end{array}
\right] _{q}:=\ln_{q}n!_{q}-\ln_{q}n_{1}!_{q}-\cdots-\ln_{q}n_{k}!_{q},
\label{q-logarithm of the q-multinomial coefficient}
\end{equation}
where 
\begin{equation}
n=\sum\limits_{i=1}^{k}n_{i},\quad n_{i}\in\mathbb{N\,}\left( i=1,\cdots
,k\right) .   \label{constraint1}
\end{equation}
Note these definitions (\ref{q-logarithm of the q-factorial}) and (\ref%
{q-logarithm of the q-multinomial coefficient}) hold for any
natural number $n\in\mathbb{N}$.

Then, we apply the $q$-Stirling's formula in (\ref%
{q-logarithm of the q-multinomial coefficient}) which uniquely leads to%
\begin{equation}
\ln_{q}\left[ 
\begin{array}{ccc}
& n &  \\ 
n_{1} & \cdots & n_{k}%
\end{array}
\right] _{q}\simeq\left\{ 
\begin{array}{ll}
\dfrac{n^{2-q}}{2-q}\cdot S_{2-q}^{\text{Tsallis}}\left( \dfrac{n_{1}}{n}%
,\cdots,\dfrac{n_{k}}{n}\right) & \text{\qquad}\left( q\neq2\right) \\ 
-S_{1}^{\text{Tsallis}}\left( n\right) +\sum\limits_{i=1}^{k}S_{1}^{\text{%
Tsallis}}\left( n_{i}\right) & \text{\qquad}\left( q=2\right),%
\end{array}
\right.   \label{one-to-one_Tsallis}
\end{equation}
where $S_{q}^{\text{Tsallis}}$ is Tsallis entropy \cite{Ts88} defined by 
$
S_{q}^{\text{Tsallis}}:=\left(1-\sum_{i=1}^{k}{p_{i}^{q}}\right)/\left(q-1\right)
\label{Tsallis entropy}
$
and $S_{1}^{\text{Tsallis}}\left( n\right) :=\ln n$. This is a
straightforward derivation of Tsallis entropy from the fundamental nonlinear
differential equation (\ref{nlde}).


\subsection{Reformulation of $q$-Gaussian distribution with scale invariance}

There are several important probability distributions associated with
Tsallis entropy such as a $q$-canonical distribution and a $q$-Gaussian
distribution. 
In this section, we derive the $q$-logarithm representation of the 
$q$-Gaussian distribution for unified studies with a fixed scale
unit of measurement.
There are several ways to derive a $q$-Gaussian distribution \cite{Ts09}.
The simplest way is the Maximum Likelihood Principle (MLP for short)
 \cite{Suyari04-LawofError}.
In the course of the derivation of $q$-Gaussian distribution in the MLP,
$q$-logarithm representation including a scaling factor $C$ is 
naturally appears.

Here $n$ observed values $x_{1}^{\prime},x_{2}^{\prime},\cdots,x_{n}^{\prime
}\in\mathbb{R}$ are given, but these values do not have the same scale unit.
Instead, there exist 
\begin{equation}
x_{1},x_{2},\cdots,x_{n}\in\mathbb{R}
\end{equation}
with a same scale unit. Each $x_{i}\in\mathbb{R}$ corresponds to
each $x_{i}^{\prime}\in\mathbb{R}$ $\left( i=1,\cdots,n\right) $,
respectively (e.g., (\ref{def of x_prime})). The $q$-logarithm likelihood
function $\log_{q}L_{q}\left( \theta\right) $ is defined by%
\begin{equation}
\log_{q}L_{q}\left( \theta\right) :=\sum\limits_{i=1}^{n}\log_{q}f\left(
x_{i}-\theta\right),   \label{q-logarithm likelihood function}
\end{equation}
where $\theta$ is a variable for this function $L_{q}$ and $f$ is a
probability density function with $x_{i}-\theta$ as a value of its
corresponding random variable.

If the function $\log_{q}L_{q}\left( \theta\right) $ of $\theta$ for any
fixed $x_{1},x_{2},\cdots,x_{n}$ attains the maximum value at%
\begin{equation}
\theta=\theta^{\ast}:=\frac{x_{1}+x_{2}+\cdots+x_{n}}{n}, 
\label{theta-star}
\end{equation}
the probability density function $f$ must be a $q$-Gaussian: 
\begin{equation}
f\left( e\right) =\frac{\exp_{q}\left( -\beta_{q}e^{2}\right) }{\int
\exp_{q}\left( -\beta_{q}e^{2}\right) de},   \label{q-Gaussian}
\end{equation}
where $\beta_{q}$ is a $q$-dependent positive constant.

See \cite{Suyari04-LawofError} for the detailed proof. Note that the
requirement (\ref{theta-star}) means that the scale units of $x_{i}\left(
i=1,\cdots ,n\right) $ should be the same among them so that this addition can
be computed.

In the course of the proof \cite{Suyari04-LawofError}, the following differential equation is derived
from the requirement of the theorem.
\begin{equation}
\frac{f^{\prime}\left( e\right) }{\left( f\left( e\right) \right) ^{q}}=a_{q}e,
\label{q-bibun_houteishiki}
\end{equation}
where $a_{q}\in\mathbb{R}$. 
Equation (\ref{q-bibun_houteishiki}) can be integrated
with respect to $e$:%
\begin{equation}
\ln_{q}f\left( e\right) =\frac{a_{q}e^{2}}{2}+C_{q}, 
\label{lnq of density fe}
\end{equation}
where $C_{q}$ is a $q$-dependent integration constant. This expression (\ref%
{lnq of density fe}) is obviously $q$-logarithm representation. If $1+\left(
1-q\right) \left( {a_{q}e^{2}}/{2}+C_{q}\right) >0$, $1+\left(
1-q\right) C_{q}>0$, then we obtain a $q$-Gaussian probability density
function (\ref{q-Gaussian}) with 
$
\beta_{q}:={-a_{q}}/
$
$\left(2\left( 1+\left( 1-q\right) C_{q}\right)\right) >0.
$
Within constraints on $C_{q}$, the arbitrariness of an integration
constant $C_{q}$ still remains.

Note that the final expression (\ref{q-Gaussian}) is clearly $q$%
-exponential representation and in this expression, $C_{q}$ is included in
both denominator and numerator of (\ref{q-Gaussian}).

In order to see a rescaling effect in the final expression (\ref{q-Gaussian}%
), the corresponding frequency distribution can be obtained as follows.

Let $\gamma_{q}$ be defined by $\gamma_{q}:=-{a_{q}}/{2}$. 
Then (\ref{lnq of density fe}) is rewritten as%
\begin{equation}
\ln_{q}f\left( e\right) =-\gamma_{q}e^{2}+C_{q}.
\end{equation}
Hence, we obtain%
\begin{equation}
\frac{f\left( e\right) }{c}=\exp_{q}\left( -\gamma_{q}\left( \frac {e}{c^{%
\frac{1-q}{2}}}\right) ^{2}\right),   \label{q-frequency dist.}
\end{equation}
where $c:=\exp_{q}\left( C_{q}\right) >0$. $f\left( e\right) $ is the
probability density function, so the left side 
${f\left( e\right) }/{c}$ is no longer a probability density function. 
But $\left({f\left( e\right) }/{c}\right)\Delta e$ represents 
frequency distribution which has scale invariance due
to arbitrariness of $c$. Obviously, under the rescaling: 
\begin{equation}
\tilde{f}\left( e\right) :=\frac{f\left( e\right) }{c},\quad\tilde {e}:=%
\frac{e}{c^{\frac{1-q}{2}}},   \label{q-gauss-rescaling}
\end{equation}
(\ref{q-frequency dist.}) is rewritten as%
\begin{equation}
\tilde{f}\left( e\right) =\exp_{q}\left( -\gamma_{q} \tilde {e}
^{2}\right) .
\end{equation}
This also represents invariance of the frequency distribution (\ref%
{q-frequency dist.}) under the rescaling (\ref{q-gauss-rescaling})
on both $e$-axis and $f\left( e\right)$-axis.

Here, for simplicity and easy understanding, we set
$y:=f\left(  e\right)  ,x:=e$.
The graphs of $\dfrac{y}{c}=\exp_{q}\left(
-\left(  \dfrac{x}{c^{\frac{1-q}{2}}}\right)  ^{2}\right)  $ for
$c=1,10,100$ and $q=1.7$ are described in Fig.3.

\begin{figure}[h]
\begin{center}
\includegraphics[scale=0.8]{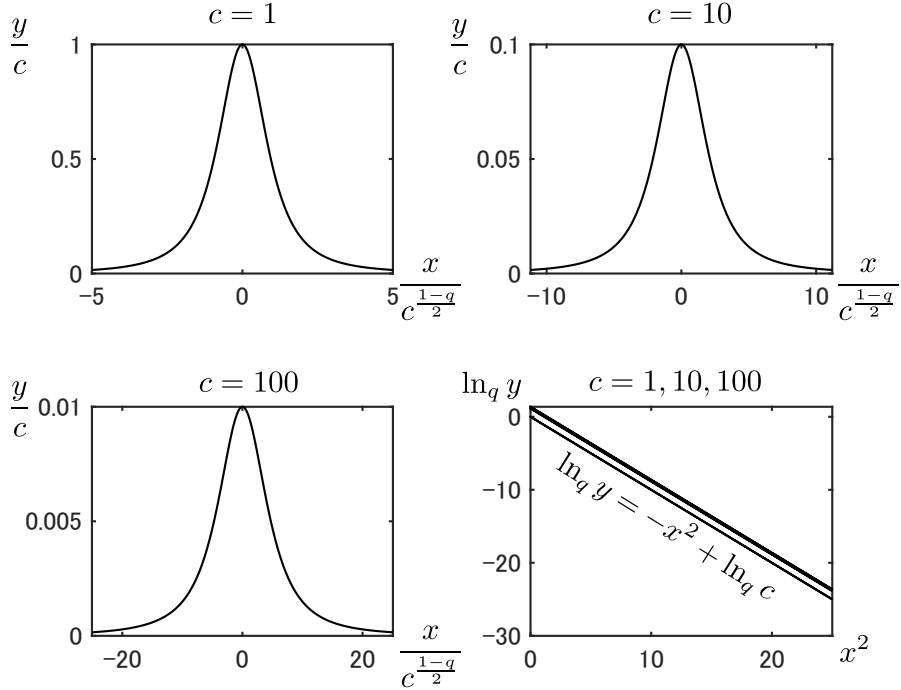}%
\caption{Graphs of $\dfrac{y}{c}=\exp_{q}\left(  -\left(  \frac{x}%
{c^{\frac{1-q}{2}}}\right)  ^{2}\right)  $ for $c=1,10,100$
 (the three graphs except the bottom right one) 
and $\ln_{q}y=-x^2+\ln_{q}c$ (the bottom right graph)
where $q=1.7$.}
\end{center}
\end{figure}

The shapes of the graphs for $c=1,10,100$ in Fig.3
are completely the same, but each scale unit of measurement is
different with each other due to the rescaling $c$.
This means under the rescaling
(\ref{q-gauss-rescaling}) the graph of $y=\exp_{q}\left(  -x^{2}\right)  $ is
invariant.
The distribution (\ref{q-frequency dist.}) can be easily transformed
into a probability distribution by imposing a normalization 
depending on each scale.

Note that when $q=1$ the scale unit on $x$-axis is \textit{fixed}
for any cases (see also (\ref{q-gauss-rescaling})) and a scaling in
(\ref{q-gauss-rescaling}) is appeared on $y$\textit{-axis only}, which is
applied to normalization in probability distributions.
In other words, when $q\neq1$, normalization should be very careful, because usual normalization
depends on scale unit on $x$-axis \cite{MSW2019}.


\section{Advantages of $q$-logarithm representation over $q$-exponential
representation through a concrete example}

In the previous two sections, $q$-exponential representation 
and $q$-logarithm representation have different purposes of expressing.
The former is for rescaling, the latter for unified understanding with a fixed scale unit.
However, for the theoretical studies including computer simulations,
$q$-logarithm representation has some crucial advantages over $q$-exponential
representation.
In particular, the non-uniqueness problems in $q$-exponential representation
is always appeared in a formulation of a probability distribution.
Through the following general example, the non-uniqueness is concretely
shown. For simplicity and ease of understanding, we present the case of a
discrete distribution. The case of a continuous distribution is similarly
discussed. After this example, the solution for this non-uniqueness problem
is given by the $q$-logarithm representation as a unique expression.

Consider the following situation such that a frequency $n_{i}\in\mathbb{N}$
of data $x_{i}$ is given by 
\begin{equation}
n_{i}=\exp_{q}\left( -x_{i}+c\right) ,\quad\left( i=1,\cdots ,k\right),
   \label{original_q-dist}
\end{equation}
where $c$ is a constant. Let the total frequency $n$ be defined
by $n:=\sum_{i=1}^{k}n_{i}.$
Then,%
\begin{equation}
n=\sum_{i=1}^{k}n_{i}=\sum_{i=1}^{k}\exp_{q}\left( -x_{i}+c\right) . 
\label{n_as_sum}
\end{equation}
We want to find a probability distribution $\left\{ p_{i}\right\} $ for
these data, so we can compute%
\begin{equation}
p_{i}:=\frac{n_{i}}{n}=\frac{n_{i}}{\sum_{i=1}^{k}n_{i}}=\frac{%
\exp_{q}\left( -x_{i}+c\right) }{\sum_{i=1}^{k}\exp_{q}\left(
-x_{i}+c\right) }.   \label{probability_q-form}
\end{equation}
When $q=1$, 
\begin{equation}
p_{i}=\frac{\exp\left( -x_{i}\right) }{\sum_{i=1}^{k}\exp\left(
-x_{i}\right) }
\end{equation}
which does not depend on $c$ and is the \textit{unique} expression using 
the only observed value $x_{i}$.
However, when $q\neq1$, innumerably many
equivalent representations for probability distribution 
(\ref{probability_q-form}) can be acceptable.
For example, for the case 
$c=c_{1}+c_{2}$ $\left( c_{1}\neq c_{2}\right) $ we have%
\begin{equation}
p_{i}=\frac{\exp_{q}\left( -x_{i}+c_{1}+c_{2}\right) }{\sum_{i=1}^{k}%
\exp_{q}\left( -x_{i}+c_{1}+c_{2}\right) }.
\end{equation}
We rewrite $\exp_{q}\left( -x_{i}+c_{1}+c_{2}\right)$
in the two kinds of representations
\begin{align}
&\exp_{q}\left( -x_{i}+c_{1}+c_{2}\right)\nonumber \\
&=\exp_{q}\left( c_{1}\right)
\exp_{q}\left( \frac{-x_{i}+c_{2}}{\left( \exp_{q}\left( c_{1}\right)
\right) ^{1-q}}\right)
=\exp_{q}\left( c_{2}\right) \exp_{q}\left( \frac{-x_{i}+c_{1}}{\left(
\exp_{q}\left( c_{2}\right) \right) ^{1-q}}\right).   \label{latter}
\end{align}
Therefore, $p_{i}$ in (\ref{probability_q-form}) is given in the two possible ways:
\begin{equation}
p_{i}=\frac{\exp_{q}\left( \frac{-x_{i}+c_{2}}{\left( \exp_{q}\left(
c_{1}\right) \right) ^{1-q}}\right) }{\sum\limits_{i=1}^{k}\exp_{q}\left( 
\frac{-x_{i}+c_{2}}{\left( \exp_{q}\left( c_{1}\right) \right) ^{1-q}}%
\right) }  
=\frac{\exp_{q}\left( \frac{-x_{i}+c_{1}}{\left( \exp_{q}\left(
c_{2}\right) \right) ^{1-q}}\right) }{\sum\limits_{i=1}^{k}\exp_{q}\left( 
\frac{-x_{i}+c_{1}}{\left( \exp_{q}\left( c_{2}\right) \right) ^{1-q}}%
\right) }.
\label{q_representation}
\end{equation}
Of course, innumerably many choices of $c_{1}$ to satisfy $c=c_{1}+c_{2}$
are available. Even for the simple representation (\ref{probability_q-form}),
there exist very many equivalent representations of a probability
distribution. This is due to \textit{arbitrary} selection of \textit{%
rescaling} and \textit{shift} for the observed values (see (\ref%
{q_representation})). These non-unique representations
such as (\ref{q_representation}) comes from the fact
that the nonlinear system (\ref{nlde}) is invariant for 
any rescaling and shift of observed values $x_{i}$.

Therefore, $q$-exponential representation as probability distribution is 
\textit{not} unique, in general. In order to avoid the non-uniqueness of $q$%
-exponential representation, $q$-logarithm representation should be used for
probability distribution. From (\ref{original_q-dist}),%
\begin{equation}
\ln_{q}n_{i}=-x_{i}+c.
\end{equation}
Hence, after some computations, we obtain 
\begin{equation}
\ln_{q}p_{i}=-{n}^{q-1}x_{i}+\left({n}^{q-1}c-\ln_{2-q}{n}\right),
\label{q-logarithm representation}
\end{equation}
where we used%
\begin{equation}
\ln_{q}\frac{y}{x}=x^{q-1}\left( \ln_{q}y-\ln_{q}x\right) .
\end{equation}
The $q$-logarithm representation (\ref{q-logarithm representation}) is 
obviously unique except for $c$.
For example, in case $c=c_{1}+c_{2}$ as stated above, the expression (\ref{q-logarithm
representation}) is invariant.

Therefore, $q$-logarithm representation should be used for probability
distribution instead of $q$-exponential representation in order to avoid
non-uniqueness.
Recently, this non-uniqueness problem is also discussed in \cite{MSW2019}
from the information geometrical points of view.


\section{Conclusion}

Long range correlations and past- or history- dependence have been studied
for many years in both linear and nonlinear systems \cite{THK18}.
In this paper, from the sense of the scale unit of measurement,
we analytically discuss how each observed data in a nonlinear system
has received influence on scale from other data
on the simplest model determined by the fundamental 
nonlinear differential equation (\ref{nlde}).
Any correlation among observed data on the dynamics (\ref{nlde})
is purely due to rescaling by the previous data, 
which yields different scale unit of measurement.
This rescaling is found to be equivalent to shift in the argument of 
the dynamics (\ref{nlde}).
These effects such as rescaling and shift result in long range correlations 
among the data.
In order to avoid different scale units on data, a corresponding logarithm
(e.g., $q$-logarithm) representation is shown to have some crucial advantages
such as uniqueness over a corresponding exponential representation.
These results can be applied to many studies in nonlinear systems.

\section*{Acknowledgement}
The first author is grateful to his son for asking the author to solve 
the arithmetic problem about the two rulers with different scale units,
which inspired the first author to find the present idea.
The first author is also grateful to Jan Korbel and Atsumi Ohara
for his careful reading and comments on the first draft.
This work was supported by JSPS KAKENHI Grant Number 17K19957 in Japan.

%

\end{document}